\documentclass[amsmath,aps,showpacs,a4paper,10pt]{revtex4}
 \usepackage{epsf}
 \usepackage{graphicx}    % Include figure files

\begin{document}

 \newcommand{\be}[1]{\begin{equation}\label{#1}}
 \newcommand{\ee}{\end{equation}}
 \newcommand{\bea}{\begin{eqnarray}}
 \newcommand{\eea}{\end{eqnarray}}
 \def\disp{\displaystyle}

 \def\gsim{ \lower .75ex \hbox{$\sim$} \llap{\raise .27ex \hbox{$>$}} }
 \def\lsim{ \lower .75ex \hbox{$\sim$} \llap{\raise .27ex \hbox{$<$}} }

 \begin{titlepage}

 \begin{flushright}
 arXiv:1706.04063
 \end{flushright}

 \title{\Large \bf Observational Constraints on Varying Alpha
 in~$\Lambda(\alpha)$CDM~Cosmology}

 \author{Hao~Wei\,}
 \email[\,email address:\ ]{haowei@bit.edu.cn}
 \affiliation{School of Physics,
 Beijing Institute of Technology, Beijing 100081, China}

 \author{Dong-Ze~Xue\,}
 \affiliation{School of Physics,
 Beijing Institute of Technology, Beijing 100081, China}

 \begin{abstract}\vspace{1cm}
 \centerline{\bf ABSTRACT}\vspace{2mm}
 In this work, we consider the so-called $\Lambda(\alpha)$CDM
 cosmology with $\Lambda\propto\alpha^{-6}$ while
 the fine-structure ``constant'' $\alpha$ is varying. In
 this scenario, the accelerated expansion of the
 universe is driven by the cosmological ``constant'' $\Lambda$
 (equivalently the vacuum energy), and the varying $\alpha$ is
 driven by a subdominant scalar field $\phi$ coupling with the
 electromagnetic field. The observational constraints on the
 varying $\alpha$ and $\Lambda\propto\alpha^{-6}$ models with
 various couplings $B_F(\phi)$ between the subdominant scalar
 field $\phi$ and the electromagnetic field are considered.
 \end{abstract}

% https://publishing.aip.org/publishing/pacs/pacs-2010-regular-edition

 \maketitle

 \end{titlepage}

 \renewcommand{\baselinestretch}{1.0}

%============================= section 1 ===================================

\section{Introduction}\label{sec1}

The year of 1998 is amazing in some sense. In this year,
 the accelerated expansion of the universe was firstly
 discovered from the observation of distant type Ia supernovae
 (SNIa)~\cite{r1}. Later, this amazing discovery was further
 confirmed by the observations of cosmic microwave background
 (CMB)~\cite{r2} and large-scale structure (LSS)~\cite{r3}
 (including baryon acoustic oscillation (BAO)~\cite{r4}
 especially). This mysterious phenomenon has formed a big challenge
 to physicists and cosmologists. It hints the existence of dark
 energy (a new component with negative pressure). The simplest
 candidate of dark energy is a tiny positive cosmological
 constant $\Lambda$ (equivalently the vacuum energy). However,
 it is hard to understand why the observed vacuum energy density is
 about 120 orders of magnitude smaller than its natural expectation
 (namely the Planck energy density). This is the so-called
 cosmological constant problem~\cite{r5,r6,r45}.

In the literature, many attempts have been made to solve (at
 least alleviate) the cosmological constant problem. Among
 them, an interesting idea is the so-called
 axiomatic approach~\cite{r7}. Based on four natural and simple
 axioms in close analogy to the Khinchin axioms (which can uniquely
 derive the Shannon entropy in information theory~\cite{r8}),
 Beck~\cite{r7} derived an explicit form for the cosmological
 constant, i.e.
 \be{eq1}
 \Lambda=
 \frac{G^2}{\hbar^4}\left(\frac{m_e}{\alpha}\right)^6\,,
 \ee
 in which $\alpha$ is the electromagnetic fine-structure
 constant, $m_e$ is the electron mass, $G$ is the gravitational
 constant, $\hbar$ is the reduced Planck constant. Accordingly,
 the vacuum energy density reads~\cite{r7}
 \be{eq2}
 \rho_\Lambda\equiv\frac{c^4\Lambda}{8\pi G}
 =\frac{Gc^4}{8\pi\hbar^4}\left(\frac{m_e}{\alpha}\right)^6\,,
 \ee
 where $c$ is the speed of light. Numerically, it gives
 $\rho_\Lambda\simeq 4.0961\,{\rm GeV/m^3}$, which can easily
 pass all the current observational constraints. We refer
 to~\cite{r7} for the detailed derivations. Note
 that Eq.~(\ref{eq1}) can also be derived in other completely
 independent approaches (see e.g.~\cite{r9,r46,r10}). We
 refer to~\cite{r23} for a brief review of these approaches.

Coincidentally, in the same year 1998, another amazing
 discovery was claimed. From the observation of distant
 quasars, Webb~{\it et al.}~\cite{r11} announced the first hint
 for the varying fine-structure ``constant'' $\alpha$. While
 the relevant observational data are accumulating~\cite{r12,
 r13,r14}, a time-varying $\alpha$ has been extensively
 considered in the literature (see e.g.~\cite{r15,r16,r17,r18,
 r19,r20,r21,r22,r23,r30,r31,r32} and references therein).
 Noting $\Lambda\propto\alpha^{-6}$ from Eq.~(\ref{eq1}), a
 time-varying $\Lambda$ follows a time-varying $\alpha$. In
 such a way, these two amazing discoveries in the year 1998 are
 dramatically connected. Actually, the cosmological implications of
 this insight have been discussed in~\cite{r23}. However,
 in~\cite{r23}, the varying $\Lambda$ and $\alpha$ were studied
 only in a phenomenological manner, but the mechanism to drive
 the varying $\alpha$ was not discussed. Therefore, we try to
 extend the work of~\cite{r23} from another perspective in
 the present work.

As is well known, the possible variations of the fundamental
 constants were firstly proposed by Dirac~\cite{r24} and
 Eddington~\cite{r25} from their large number hypothesis.
 The most observationally sensitive one is the electromagnetic
 fine-structure ``constant'' $\alpha\equiv e^2/(\hbar c)$. A
 varying $\alpha$ might be due to a varying speed of light
 $c$~\cite{r26}, while Lorentz invariance is broken. Another
 possibility for a varying $\alpha$ is due to a varying
 electron charge $e$, which was firstly proposed
 by Bekenstein~\cite{r27} in 1982, while local gauge and
 Lorentz invariance are preserved. This is a dilaton theory
 with coupling to the electromagnetic $F^2$ part of the
 Lagrangian, but not to the other gauge fields. It has been
 generalized to a so-called BSBM model~\cite{r28,r29} after the
 first observational hint of varying $\alpha$ from the quasar
 absorption spectra in 1998~\cite{r11}. In fact, the main spirit of
 Bekenstein-type models is using a scalar field $\phi$ coupling
 with the electromagnetic field to drive the varying $\alpha$.

In the literature, there exist two main types of varying
 $\alpha$ models driven by a scalar field $\phi$, depending
 on the role played by $\phi$. The first one is using the
 scalar field $\phi$ to simultaneously drive the accelerated
 expansion of the universe and the varying $\alpha$ (see
 e.g.~\cite{r20,r21,r18}). That is, the scalar field $\phi$
 also plays the role of dark energy, and it is the dominant
 component of the universe. On the contrary, the second one is using
 the scalar field $\phi$ to drive only the varying $\alpha$.
 The accelerated expansion of the universe is instead driven by the
 cosmological constant $\Lambda$ (equivalently the vacuum
 energy), while the scalar field $\phi$ is subdominant and its
 only role is to drive the varying $\alpha$ (see
 e.g.~\cite{r19,r28,r29}). In the present work, we adopt the
 second perspective naturally. For simplicity, we consider a
 subdominant quintessence with a canonical kinetic energy
 as the simplest scalar field $\phi$ to drive the varying $\alpha$.

The rest of this paper is organized as followings. In
 Sec.~\ref{sec2}, we setup the varying $\alpha$ model driven by
 a subdominant quintessence $\phi$ in $\Lambda(\alpha)$CDM cosmology
 with $\Lambda\propto\alpha^{-6}$. Then, in Sec.~\ref{sec3}, we
 consider the observational constraints on the varying $\alpha$ and
 $\Lambda\propto\alpha^{-6}$ models with various couplings
 $B_F(\phi)$. The brief concluding remarks are given in
 Sec.~\ref{sec4}.

%============================= Fig. 1 =================================

 \begin{center}
 \begin{figure}[htb]
 \centering
 \includegraphics[width=0.43\textwidth]{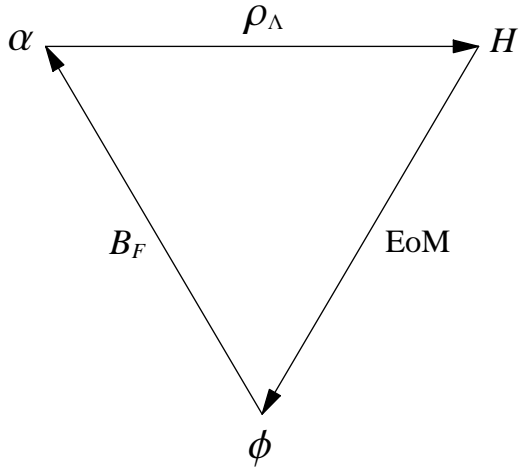}
 \caption{\label{fig1} The relations between the scalar field
 $\phi$, the varying fine-structure ``constant'' $\alpha$,
 and the Hubble parameter $H$. See the text for details.}
 \end{figure}
 \end{center}

%======================================================================

\vspace{-12mm}  % used here just for a comfortable typesetting

%============================= section 2 ===================================

\section{Varying alpha driven by quintessence in
 $\Lambda(\alpha)$CDM cosmology}\label{sec2}

Following e.g.~\cite{r20,r21,r30,r31}, the
 relevant action reads
 \be{eq3}
 {\cal S}=-\frac{m_p^2}{2}\int d^4 x\,\sqrt{-g}\left(R-2\Lambda
 \right)+\int d^4 x\,\sqrt{-g}\,{\cal L}_\phi-\frac{1}{4}\int d^4 x
 \,\sqrt{-g}\,B_F(\phi)\,F_{\mu\nu}F^{\mu\nu}+{\cal S}_m\,,
 \ee
 where $R$ is the Ricci scalar; $g$ is the determinant of the
 metric $g_{\mu\nu}$; $m_p\equiv (8\pi G)^{-1/2}$ is the
 reduced Planck mass; $F_{\mu\nu}$ are the components of the
 electromagnetic field tensor; ${\cal S}_m$ is the action of
 pressureless matter; we have set the units $\hbar=c=1$; we
 can safely ignore the contribution of radiation;
 ${\cal L}_\phi$ is the Lagrangian of the subdominant scalar
 field $\phi$. For a subdominant quintessence, it is given by
 \be{eq4}
 {\cal L}_\phi=
 \frac{1}{2}\partial_\mu \phi\partial^\mu \phi-V(\phi)\,,
 \ee
 where $V(\phi)$ is the potential. Noting that the coupling
 $B_F$ takes the place of $e^{-2}$
 in Eq.~(\ref{eq3}) actually~\cite{r32}, one can easily see
 that the effective fine-structure ``constant''
 is given by~\cite{r20,r21,r27,r30}
 \be{eq5}
 \alpha=\frac{\alpha_0}{B_F(\phi)}\,,
 \ee
 and then
 \be{eq6}
 \frac{\Delta\alpha}{\alpha}\equiv\frac{\alpha-\alpha_0}{\alpha_0}=
 B_F^{-1}(\phi)-1\,,
 \ee
 where the subscript ``0'' indicates the present value of
 corresponding quantity. By definition, the present value of
 $B_F$ should be equal to $1$. In general, $\phi$ and hence
 $\alpha$ are functions of spacetime. However, as is well
 known, we can safely ignore their spatial variation, and only
 consider the homogeneous $\phi$ and $\alpha$ in the present
 work. Throughout, we assume that only the electron charge
 $e$ is varying, and all the other fundamental constants
 $\hbar$, $G$, $c$, $m_e$ are true constants.
 Thus, $\rho_\Lambda\propto\Lambda\propto\alpha^{-6}$.
 Using Eq.~(\ref{eq5}), we get
 \be{eq7}
 \rho_\Lambda=\rho_{\Lambda 0}\left(\frac{\alpha}{\alpha_0}
 \right)^{-6}=\rho_{\Lambda 0}B_F^6(\phi)\,.
 \ee
 Considering a flat Friedmann-Robertson-Walker (FRW) universe,
 the corresponding Friedmann equation and Raychaudhuri
 equation are given by
 \bea
 &\disp H^2=\frac{1}{3m_p^2}\left(\rho_\Lambda+\rho_m\right)\,,
 \label{eq8}\\[0.3mm]
 &\disp\dot{H}=-\frac{1}{2m_p^2}\left(\rho_\Lambda+\rho_m+p_\Lambda
 +p_m\right)=-\frac{\rho_m}{2m_p^2}\,,\label{eq9}
 \eea
 respectively, where $H\equiv\dot{a}/a$ is
 the Hubble parameter; $a=(1+z)^{-1}$ is the scale factor (we
 have set $a_0=1$); $z$ is the redshift; a dot denotes a
 derivative with respect to the cosmic time $t$; $\rho_m$ is
 the energy density of dust matter;
 $p_\Lambda=-\rho_\Lambda$ and $p_m=0$ are the pressures of the
 vacuum energy and dust matter, respectively. We have
 safely ignored the subdominant scalar field $\phi$ and the
 electromagnetic field in Eqs.~(\ref{eq8}) and (\ref{eq9}). On
 the other hand, from the total energy conservation equation
 $\dot{\rho}_{tot}+3H(\rho_{tot}+p_{tot})=0$, we find that
 $\dot{\rho}_m+3H\rho_m=-\dot{\rho}_\Lambda\not=0$. So,
 $\rho_m$ is no longer proportional to $a^{-3}$ (see \cite{r23}
 for a detailed discussion on this issue). The equation of
 motion (EoM) for the subdominant scalar field $\phi$ reads
 \be{eq10}
 \ddot{\phi}+3H\dot{\phi}+V_{,\phi}=0\,,
 \ee
 where the subscript ``$,\phi$'' denotes the derivative with
 respect to $\phi$. In principle, there should be an additional
 term proportional to $F_{\mu\nu}F^{\mu\nu}$ and the derivative
 of $B_F$~\cite{r30} in the right hand side
 of Eq.~(\ref{eq10}), due to the coupling between the scalar
 field and the electromagnetic field. However, it could be
 safely ignored thanks to the following facts: (i) the
 derivative of $B_F$ is actually equivalent to $\dot{\alpha}$,
 which is very tiny (given equivalence principle
 constraints~\cite{r30,r31}); (ii) the statistical average of
 the term $F_{\mu\nu}F^{\mu\nu}$ over a current state of the
 universe is zero~\cite{r20}.

In Fig.~\ref{fig1}, we show the relations between the scalar
 field $\phi$, the varying fine-structure ``constant''
 $\alpha$, and the Hubble parameter $H$. The subdominant scalar
 field $\phi$ drives the varying $\alpha$ through the coupling
 $B_F$ according to Eq.~(\ref{eq5}). The varying $\alpha$
 affects the Hubble parameter $H$ (which characterizes the
 cosmic expansion) through $\rho_\Lambda\propto\alpha^{-6}$
 in Eq.~(\ref{eq8}). The Hubble parameter $H$ affects the
 evolution of the scalar field $\phi$ through the friction
 term proportional to $H$ in the EoM given by Eq.~(\ref{eq10}).

For convenience, we recast the evolution equations with
 dimensionless quantities. Substituting Eqs.~(\ref{eq9}) and
 (\ref{eq7}) into Eq.~(\ref{eq8}), we obtain
 \be{eq11}
 H^2=\frac{\rho_{\Lambda 0} B_F^6}{3m_p^2}-
 \frac{2}{3}\dot{H}\,.
 \ee
 Using the relation $\dot{f}=-(1+z)Hf^\prime$ (where a prime
 denotes a derivative with respect to the redshift $z$), we
 recast Eq.~(\ref{eq11}) as
 \be{eq12}
 E^2=\left(1-\Omega_{m0}\right)B_F^6
 +\frac{2}{3}(1+z)EE^\prime\,,
 \ee
 where $E\equiv H/H_0$
 and $\Omega_{\Lambda 0}\equiv\rho_{\Lambda 0}/(3m_p^2 H_0^2)
 =1-\rho_{m0}/(3m_p^2 H_0^2)\equiv 1-\Omega_{m0}$. Introducing
 $\hat{\varphi}\equiv\phi/m_p$ and
 $U(\hat{\varphi})\equiv V(\phi)/(m_p^2 H_0^2)$, we recast
 Eq.~(\ref{eq10}) as
 \be{eq13}
 (1+z)^2 E^2 \hat{\varphi}^{\prime\prime}+(1+z)E\left[\,(1+z)E^\prime
 -2E\,\right]\hat{\varphi}^\prime+U_{,\hat{\varphi}}=0\,.
 \ee
 For simplicity, following e.g.~\cite{r19,r28,r29}, we only
 consider the scalar field $\phi$ without potential in this
 work, and hence Eq.~(\ref{eq13}) becomes
 \be{eq14}
 (1+z) E^2 \hat{\varphi}^{\prime\prime}+E\left[\,(1+z)E^\prime-
 2E\,\right]\hat{\varphi}^\prime=0\,.
 \ee
 Once the coupling $B_F(\hat{\varphi})$ and the initial
 conditions are given, one can numerically solve the coupled
 2nd order differential equations (\ref{eq12}) and (\ref{eq14})
 to obtain $\hat{\varphi}$ and $E$ as functions of the redshift
 $z$. Then, $\Delta\alpha/\alpha$ as a function of the redshift $z$
 is on hand by using Eq.~(\ref{eq6}).

%============================= section 3 ===================================

\section{Observational constraints on the varying alpha models}\label{sec3}

One can constrain the varying $\alpha$ and
 $\Lambda\propto\alpha^{-6}$ models by using the observational
 data, if the theoretical $\Delta\alpha/\alpha$ as a function
 of the redshift $z$ is known. Here, we consider
 the observational $\Delta\alpha/\alpha$ dataset given
 in~\cite{r14,r33,r34}, which consists of 293
 usable $\Delta\alpha/\alpha$ data from the absorption systems
 in the spectra of distant quasars (note that two outliers
 should be removed from the full numerical data of 295 quasar
 absorption systems~\cite{r14,r33,r34}), over the absorption
 redshift range $0.2223\leq z_{abs}\leq 4.1798$. Note that all
 the 293 $\Delta\alpha/\alpha$ data are of ${\cal O}(10^{-5})$.
 The $\chi^2$ from these 293 $\Delta\alpha/\alpha$ data is given by
 \be{eq15}
 \chi^2_\alpha=\sum\limits_i \frac{\left[\,
 (\Delta\alpha/\alpha)_{{\rm th},i}-
 (\Delta\alpha/\alpha)_{{\rm obs},i}\,\right]^2}{\sigma_i^2}\,,
 \ee
 where $\sigma_i^2=\sigma^2_{{\rm stat},i}+\sigma^2_{{\rm rand},i}$
 (see Sec.~3.5.3 of~\cite{r14} and the instructions of~\cite{r33,r34}
 for the technical details of $\sigma_{\rm rand}$ and the error
 budget). Note that these $\Delta\alpha/\alpha$ data can
 tightly constrain the model parameters in the coupling $B_F$,
 but the constraints on the model parameter $\Omega_{m0}$ are
 too loose. Therefore, the data from the other cosmological
 observations, such as SNIa, CMB and BAO, are required to
 properly constrain the model parameter $\Omega_{m0}$. Here, we
 consider the same SNIa~\cite{r35}, CMB~\cite{r37,r38,r39,r40}
 and BAO~\cite{r4} data as in~\cite{r23}, and the corresponding
 $\chi^2$ are given with detail in Sec.~3.1 of~\cite{r23}.
 The total $\chi^2$ from the combined $\Delta\alpha/\alpha$,
 SNIa, CMB and BAO data is given by
 \be{eq16}
 \chi^2=\chi^2_\alpha+\tilde{\chi}^2_\mu+\chi^2_R+\chi^2_A\,,
 \ee
 where $\tilde{\chi}^2_\mu$, $\chi^2_R$ and $\chi^2_A$ are all
 given in Sec.~3.1 of~\cite{r23}. The best-fit model parameters are
 determined by minimizing the total $\chi^2$. As
 in~\cite{r36,r41}, the $68.3\%$ confidence level is determined
 by $\Delta\chi^2\equiv\chi^2-\chi^2_{min}\leq 1.0$, $2.3$,
 $3.53$, $4.72$ for $n_p=1$, $2$, $3$, $4$, respectively, where
 $n_p$ is the number of free model parameters. Similarly, the
 $95.4\%$ confidence level is determined by
 $\Delta\chi^2\equiv\chi^2-\chi^2_{min}\leq 4.0$, $6.18$,
 $8.02$, $9.72$ for $n_p=1$, $2$, $3$, $4$, respectively.

%============================= Fig. 2 =================================

 \begin{center}
 \begin{figure}[tb]
 \centering
 \vspace{-12mm}  % used here just for a comfortable typesetting
 \includegraphics[width=0.9\textwidth]{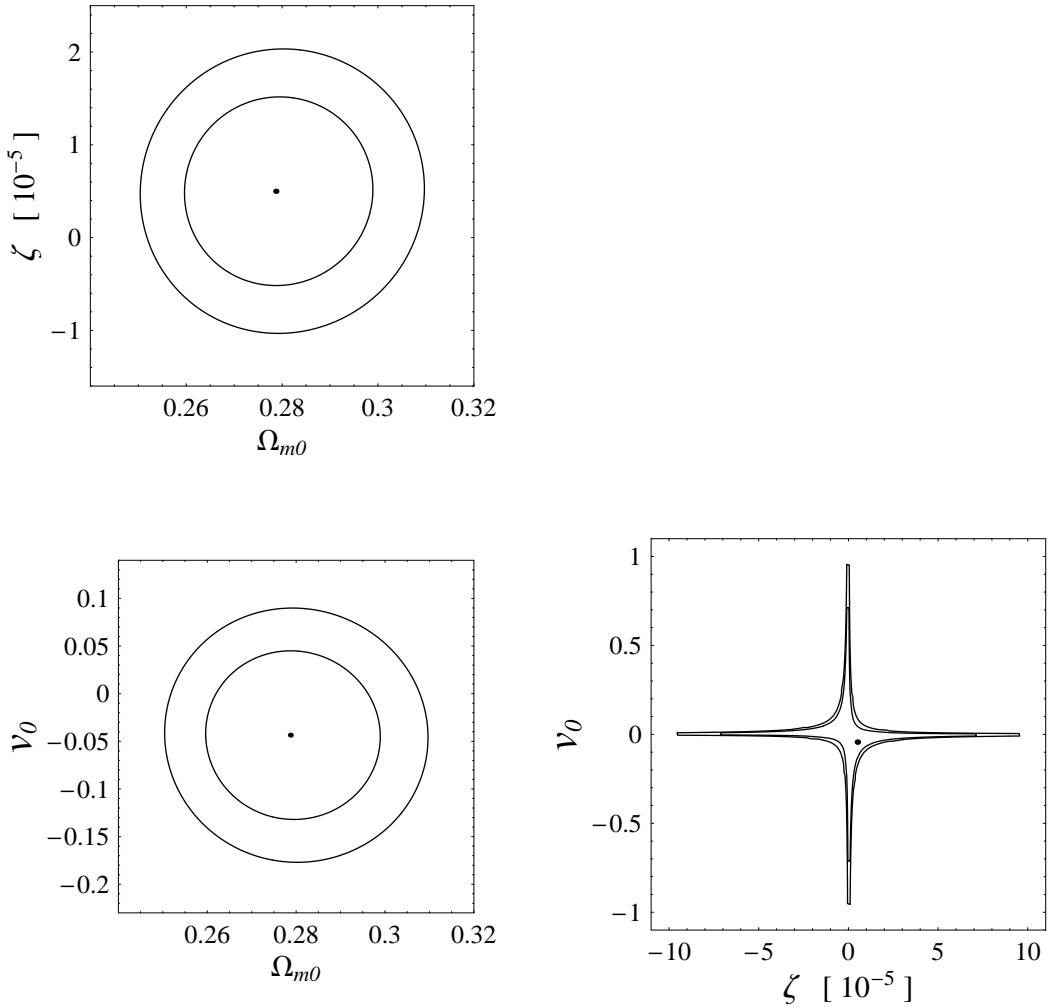}
 \caption{\label{fig2} The $68.3\%$ and $95.4\%$ confidence
 level contours in the $\Omega_{m0}-\zeta$, $\Omega_{m0}-v_0$
 and $\zeta-v_0$ planes for the case of linear coupling. Note
 that $\zeta$ is given in units of $10^{-5}$. The best-fit
 parameters are also indicated by the black solid points.}
 \vspace{-3mm} % used here just for a comfortable typesetting
 \end{figure}
 \end{center}

%======================================================================

\vspace{-15mm}  % used here just for a comfortable typesetting

%============================= section 3.1 ===================================

\subsection{Linear coupling}\label{sec3a}

At first, we consider the linear coupling~\cite{r20,r21}
 \be{eq17}
 B_F(\hat{\varphi})=1-\zeta\left(\hat{\varphi}-\hat{\varphi}_0
 \right)\,,
 \ee
 where $\zeta$ is a constant. In fact, this is the mostly
 considered coupling in the literature, since it is the
 simplest one. To simplify the initial conditions, we redefine
 $\varphi\equiv\hat{\varphi}-\hat{\varphi}_0$, and then
 $B_F(\varphi)=1-\zeta\varphi$. Now, the evolution equations
 (\ref{eq12}) and (\ref{eq14}) become
 \bea
 &\disp E^2=\left(1-\Omega_{m0}\right)B_F^6(\varphi)
 +\frac{2}{3}\left(1+z\right)EE^\prime\,,\label{eq18}\\[2mm]
 &\left(1+z\right) E^2 \varphi^{\prime\prime}+E\left[\,\left(
 1+z\right)E^\prime-2E\,\right]\varphi^\prime=0\,,\label{eq19}
 \eea
 which are the coupled 2nd order differential equations. By
 definition, the corresponding initial conditions are given by
 $E(z=0)=1$, $\varphi(z=0)=0$ and
 $\varphi^\prime(z=0)=\varphi^\prime_0=v_0$, where $v_0$ is a
 constant and will be determined by the observational data.
 In this case, there are three free model parameters, namely
 $\Omega_{m0}$, $\zeta$ and $v_0$. Note that if $\zeta=0$,
 we have $B_F\equiv 1$ and then $\alpha=const.$,
 $\Lambda\propto\alpha^{-6}=const.$, namely the model reduces
 to a constant $\alpha$ in ordinary $\Lambda$CDM cosmology. We
 can numerically solve the coupled 2nd order differential
 equations (\ref{eq18}) and (\ref{eq19}) with the initial
 conditions mentioned above to obtain $\varphi$ and $E$ as
 functions of the redshift $z$. Then, $\Delta\alpha/\alpha$
 as a function of the redshift $z$ is on hand by using
 Eq.~(\ref{eq6}). By minimizing the corresponding total
 $\chi^2$ in Eq.~(\ref{eq16}), we find the best-fit model
 parameters $\Omega_{m0}=0.2787$, $\zeta=0.4995\times 10^{-5}$,
 and $v_0=-0.0435$, while $\chi^2_{min}=869.6$
 and $\chi^2_{min}/dof=0.9972$. In Fig.~\ref{fig2}, we also
 present the corresponding $68.3\%$ and $95.4\%$ confidence
 level contours in the $\Omega_{m0}-\zeta$, $\Omega_{m0}-v_0$
 and $\zeta-v_0$ planes. From Fig.~\ref{fig2}, it is easy to
 see that $\zeta$ is tightly constrained to a narrow range of
 ${\cal O}(10^{-5})$, thanks to the 293 $\Delta\alpha/\alpha$
 data of ${\cal O}(10^{-5})$. On the other hand, $\zeta=0$ is
 within the $1\sigma$ region, and hence a constant $\alpha$ in
 ordinary $\Lambda$CDM cosmology is fully consistent with the
 observational data.

%============================= Fig. 3 =================================

 \begin{center}
 \begin{figure}[tb]
 \centering
 \vspace{-9mm}  % used here just for a comfortable typesetting
 \includegraphics[width=0.9\textwidth]{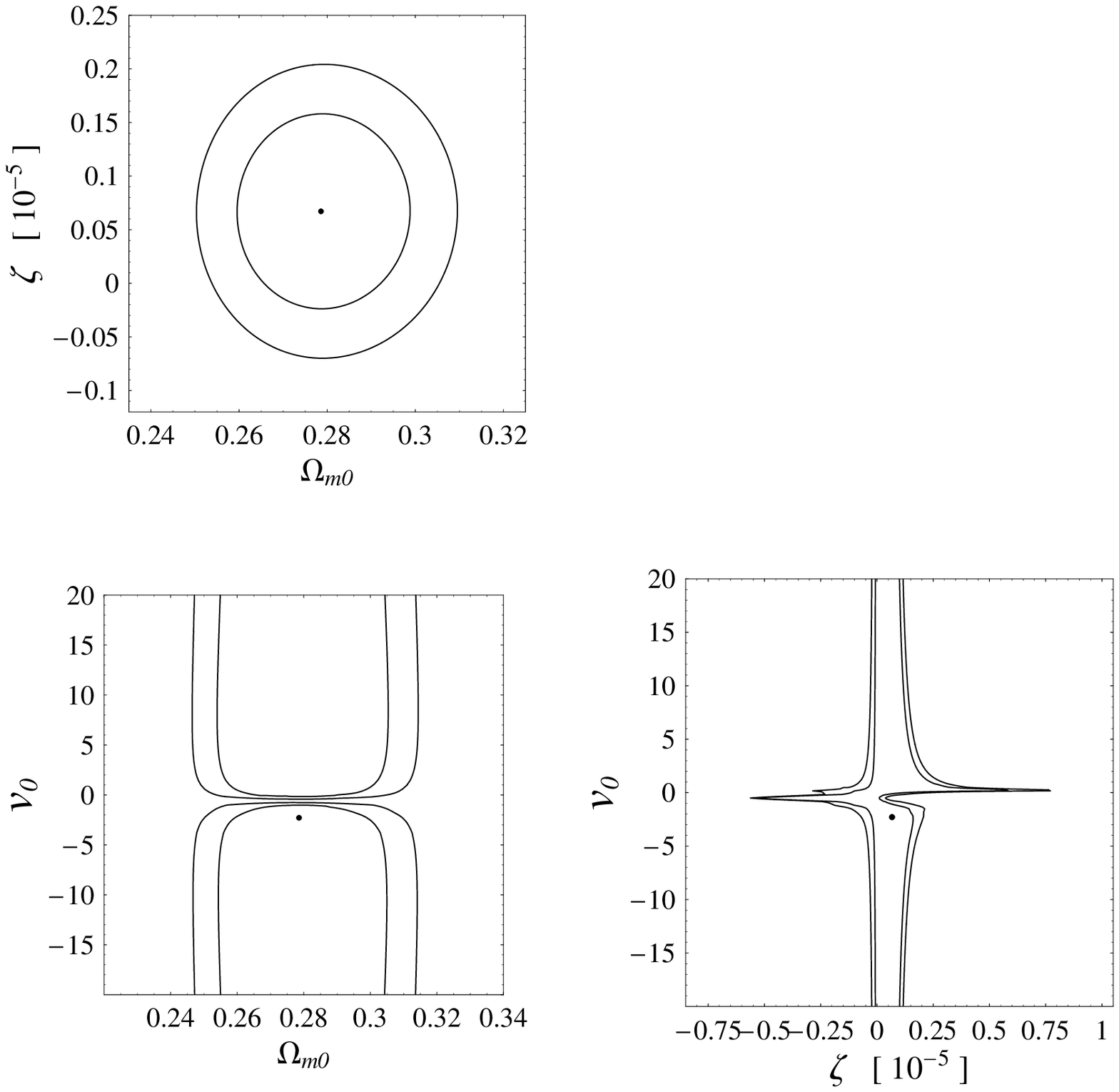}
 \caption{\label{fig3} The same as in Fig.~\ref{fig2}, except
 for the case of power-law coupling. Note that the ``gap'' in
 the $\Omega_{m0}-v_0$ plane corresponds to the ``hollow'' in
 the $\zeta-v_0$ plane.}
 \end{figure}
 \end{center}

%======================================================================

\vspace{-12mm}  % used here just for a comfortable typesetting

%============================= section 3.2 ===================================

\subsection{Power-law coupling}\label{sec3b}

Here, we consider the power-law coupling~\cite{r20,r21}
 \be{eq20}
 B_F(\hat{\varphi})=\left(\frac{\hat{\varphi}}{\hat{\varphi}_0}
 \right)^\zeta\,,
 \ee
 where $\hat{\varphi}_0\not=0$, and $\zeta$ is a constant. To
 simplify the initial conditions, we redefine
 $\varphi\equiv\hat{\varphi}/\hat{\varphi}_0$, and then
 $B_F(\varphi)=\varphi^\zeta$. However, this form is
 pathological. Noting
 $B_F(\varphi)=\varphi^\zeta=\exp(\zeta\ln\varphi)$, it will
 become complex number for $\varphi<0$. To avoid this problem,
 we should instead consider another form,
 \be{eq21}
 B_F(\varphi)=|\varphi|^\zeta\,,
 \ee
 which is equivalent to
 $B_F(\hat{\varphi})=|\hat{\varphi}/\hat{\varphi}_0|^\zeta$,
 where $|x|$ denotes the absolute value of $x$.
 With $\varphi\equiv\hat{\varphi}/\hat{\varphi}_0$, the
 evolution equations (\ref{eq12}) and (\ref{eq14}) become
 the ones given in Eqs.~(\ref{eq18}) and (\ref{eq19}). By
 definition, the corresponding initial conditions are given
 by $E(z=0)=1$, $\varphi(z=0)=1$ and
 $\varphi^\prime(z=0)=\varphi^\prime_0=v_0$, where $v_0$ is a
 constant and will be determined by the observational data.
 Note that the initial condition $\varphi(z=0)=1$ is different
 from the case of linear coupling by definition.
 In this case, there are three free model parameters, namely
 $\Omega_{m0}$, $\zeta$ and $v_0$. Note that if $\zeta=0$,
 we have $B_F\equiv 1$ and then $\alpha=const.$,
 $\Lambda\propto\alpha^{-6}=const.$, namely the model reduces
 to a constant $\alpha$ in ordinary $\Lambda$CDM cosmology. We
 can numerically solve the coupled 2nd order differential
 equations (\ref{eq18}) and (\ref{eq19}) with the initial
 conditions mentioned above to obtain $\varphi(z)$, $E(z)$,
 and then $\Delta\alpha/\alpha(z)$. By minimizing
 the corresponding total $\chi^2$ in Eq.~(\ref{eq16}), we
 find the best-fit model parameters $\Omega_{m0}=0.2786$,
 $\zeta=0.0672\times 10^{-5}$, and $v_0=-2.2871$, while
 $\chi^2_{min}=868.527$ and $\chi^2_{min}/dof=0.9960$. In
 Fig.~\ref{fig3}, we also present the corresponding $68.3\%$
 and $95.4\%$ confidence level contours in
 the $\Omega_{m0}-\zeta$, $\Omega_{m0}-v_0$
 and $\zeta-v_0$ planes. From Fig.~\ref{fig3}, it is easy to
 see that $\zeta$ is tightly constrained to a narrow range of
 ${\cal O}(10^{-6})$, thanks to the 293 $\Delta\alpha/\alpha$
 data of ${\cal O}(10^{-5})$. On the other hand, $\zeta=0$ is
 within the $1\sigma$ region, and hence a constant $\alpha$ in
 ordinary $\Lambda$CDM cosmology is fully consistent with the
 observational data. Note that the observational data cannot
 well constrain the parameter $v_0$.

%============================= Fig. 4 =================================

 \begin{center}
 \begin{figure}[tb]
 \centering
 \vspace{-15mm}  % used here just for a comfortable typesetting
 \includegraphics[width=0.9\textwidth]{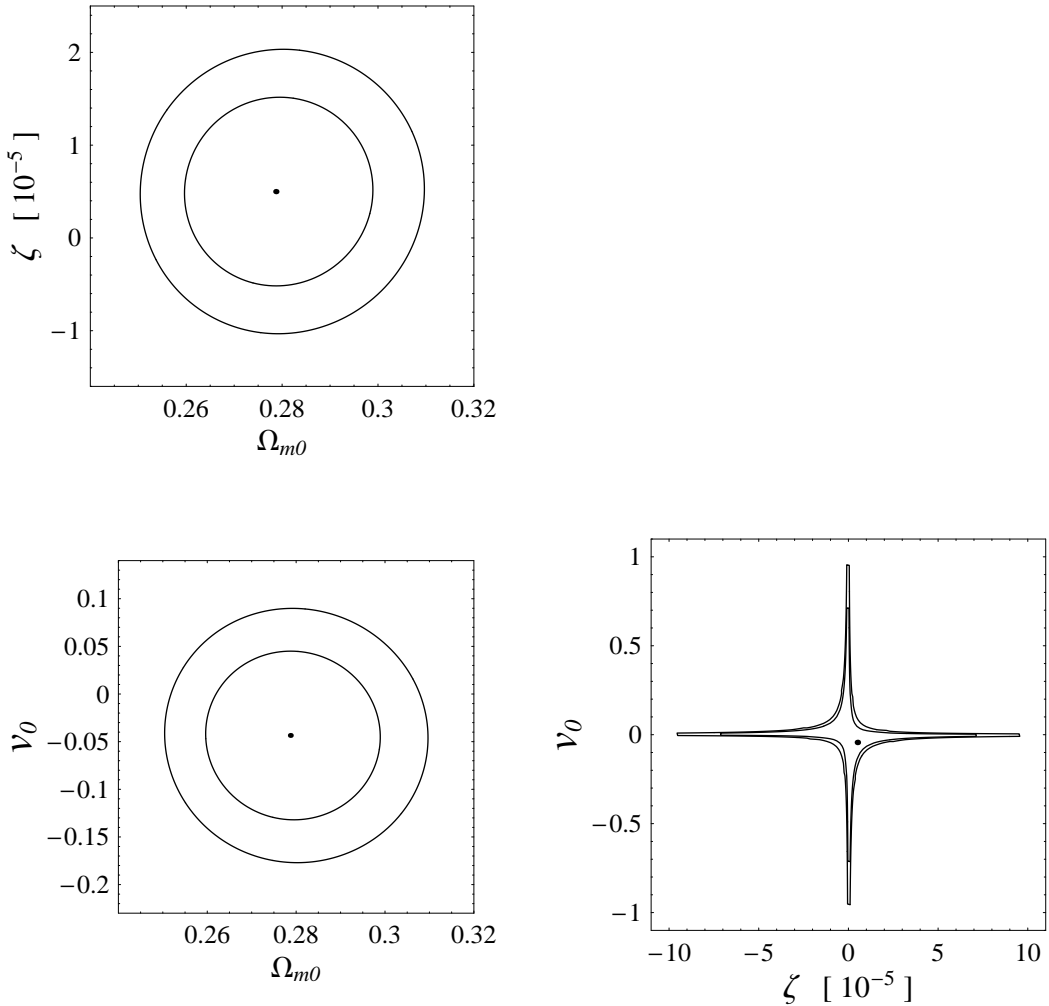}
 \caption{\label{fig4} The same as in Fig.~\ref{fig2}, except
 for the case of exponential coupling.}
 \end{figure}
 \end{center}

%======================================================================

\vspace{-12.9mm}  % used here just for a comfortable typesetting

%============================= section 3.3 ===================================

\subsection{Exponential coupling}\label{sec3c}

Let us turn to the exponential coupling~\cite{r20,r21}
 \be{eq22}
 B_F(\hat{\varphi})=\exp\left(
 -\zeta(\hat{\varphi}-\hat{\varphi}_0)\right)\,,
 \ee
 where $\zeta$ is a constant. We redefine
 $\varphi\equiv\hat{\varphi}-\hat{\varphi}_0$, and then
 $B_F(\varphi)=\exp\left(-\zeta\varphi\right)$, while the
 evolution equations (\ref{eq12}) and (\ref{eq14}) become
 the ones given in Eqs.~(\ref{eq18}) and (\ref{eq19}). By
 definition, the corresponding initial conditions are given
 by $E(z=0)=1$, $\varphi(z=0)=0$ and
 $\varphi^\prime(z=0)=\varphi^\prime_0=v_0$, where $v_0$ is a
 constant and will be determined by the observational data.
 In this case, there are three free model parameters, namely
 $\Omega_{m0}$, $\zeta$ and $v_0$. Note that if $\zeta=0$,
 we have $B_F\equiv 1$ and then $\alpha=const.$,
 $\Lambda\propto\alpha^{-6}=const.$, namely the model reduces
 to a constant $\alpha$ in ordinary $\Lambda$CDM cosmology. We
 can numerically solve the coupled 2nd order differential
 equations (\ref{eq18}) and (\ref{eq19}) with the initial
 conditions mentioned above to obtain $\varphi(z)$, $E(z)$,
 and then $\Delta\alpha/\alpha(z)$. By minimizing
 the corresponding total $\chi^2$ in Eq.~(\ref{eq16}), we
 find the best-fit model parameters $\Omega_{m0}=0.2787$,
 $\zeta=0.4994\times 10^{-5}$, and $v_0=-0.0435$, while
 $\chi^2_{min}=869.6$ and $\chi^2_{min}/dof=0.9972$. In
 Fig.~\ref{fig4}, we also present the corresponding $68.3\%$
 and $95.4\%$ confidence level contours in the
 $\Omega_{m0}-\zeta$, $\Omega_{m0}-v_0$ and $\zeta-v_0$ planes.
 It is worth noting that the best-fit model parameters and the
 contours are almost the same as in the case of linear
 coupling, while the differences are very tiny. This is not
 surprising, since $B_F(\varphi)=\exp\left(-\zeta\varphi\right)
 \simeq 1-\zeta\varphi+{\cal O}(\zeta^2)$ for $\zeta\ll 1$.

%============================= section 3.4 ===================================

\subsection{Polynomial coupling}\label{sec3d}

\begin{center}
 \begin{figure}[p]
 \centering
 \vspace{-11.5mm}  % used here just for a comfortable typesetting
 \includegraphics[width=0.9\textwidth]{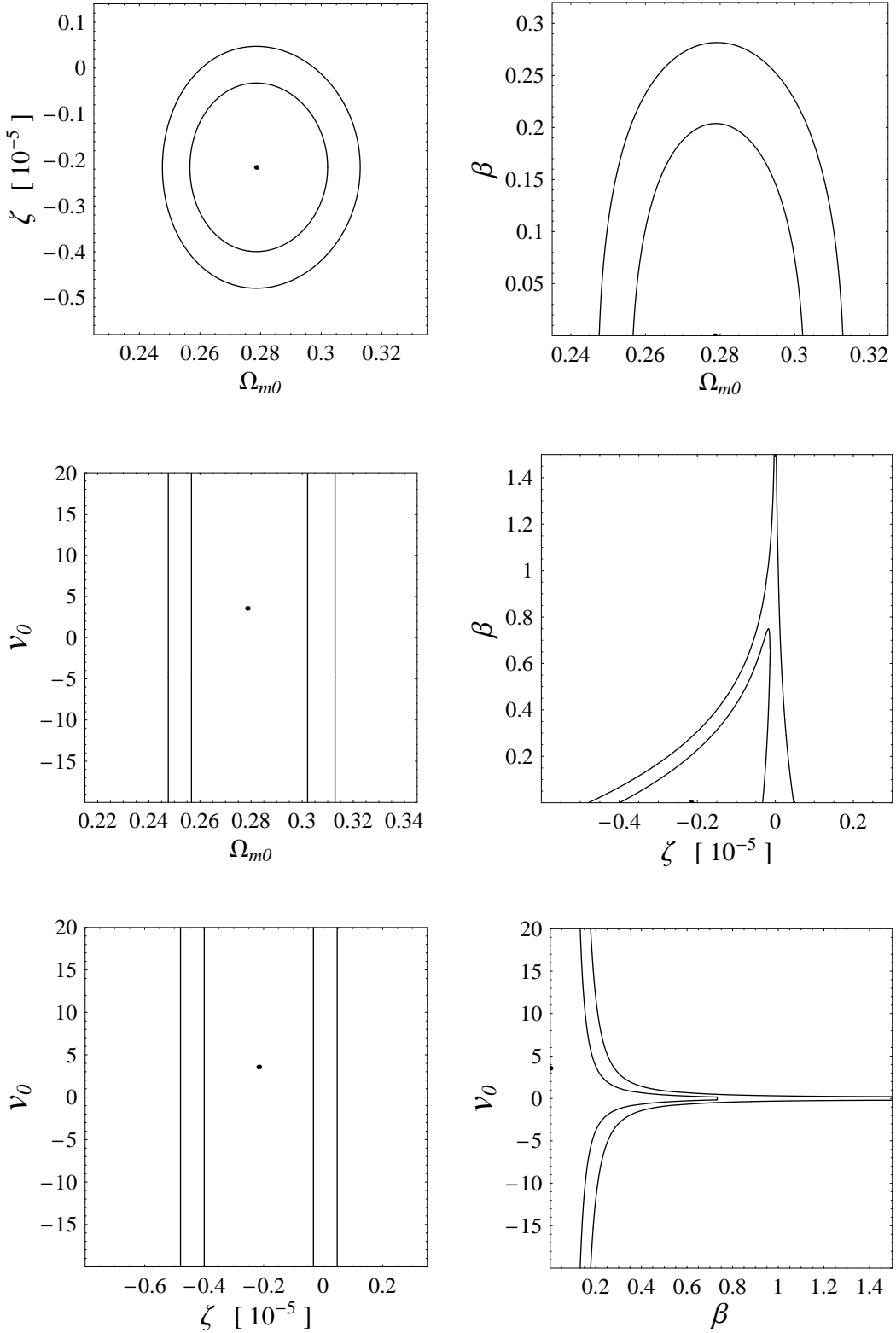}
 \caption{\label{fig5} The $68.3\%$ and $95.4\%$ confidence
 level contours in the $\Omega_{m0}-\zeta$,
 $\Omega_{m0}-\beta$, $\Omega_{m0}-v_0$, $\zeta-\beta$,
 $\zeta-v_0$ and $\beta-v_0$ planes for the case of polynomial
 coupling. Note that $\zeta$ is given in units of $10^{-5}$.
 The best-fit parameters are also indicated by the black solid
 points.}
 \end{figure}
 \end{center}

Finally, we consider the polynomial coupling~\cite{r20,r21}
 \be{eq23}
 B_F(\hat{\varphi})=1-\zeta\left(\hat{\varphi}-\hat{\varphi}_0
 \right)^\beta\,,
 \ee
 where $\zeta$ and $\beta$ are both constants. Again, we
 redefine $\varphi\equiv\hat{\varphi}-\hat{\varphi}_0$, and
 then $B_F(\varphi)=1-\zeta\varphi^\beta$. Similar to the case
 of power-law coupling, this form is pathological. Noting
 $B_F(\varphi)=1-\zeta\varphi^\beta=
 1-\zeta\exp(\beta\ln\varphi)$, it will become complex number
 for $\varphi<0$. To avoid this problem, we should instead
 consider
 \be{eq24}
 B_F(\varphi)=1-\zeta|\varphi|^\beta\,,
 \ee
 which is equivalent to
 $B_F(\hat{\varphi})=1-\zeta\left|\hat{\varphi}-\hat{\varphi}_0
 \right|^\beta$. With $\varphi\equiv\hat{\varphi}-\hat{\varphi}_0$,
 the evolution equations (\ref{eq12}) and (\ref{eq14}) become
 the ones given in Eqs.~(\ref{eq18}) and (\ref{eq19}). By
 definition, the corresponding initial conditions are given
 by $E(z=0)=1$, $\varphi(z=0)=0$ and
 $\varphi^\prime(z=0)=\varphi^\prime_0=v_0$, where $v_0$ is a
 constant and will be determined by the observational data.
 In this case, there are four free model parameters, namely
 $\Omega_{m0}$, $\zeta$, $\beta$ and $v_0$. Noting the initial
 condition $\varphi(z=0)=0$, if $\beta<0$, we find that $B_F$
 will diverge at $z=0$ (however, this is not a problem for
 the case of power-law coupling, since its corresponding
 initial condition is $\varphi(z=0)=1$, rather than $0$). On
 the other hand, if $\beta=0$ exactly, $B_F=1-\zeta\not=1$ at $z=0$
 for a non-zero $\zeta$ (note that the present value of $B_F$
 should be equal to 1 by definition). Thus, we should require
 \be{eq25}
 \beta>0\,.
 \ee
 Note that if $\zeta=0$, we have $B_F\equiv 1$ and then
 $\alpha=const.$, $\Lambda\propto\alpha^{-6}=const.$, namely
 the model reduces to a constant $\alpha$ in
 ordinary $\Lambda$CDM cosmology. We can numerically solve the
 coupled 2nd order differential equations (\ref{eq18}) and
 (\ref{eq19}) with the initial conditions mentioned above to
 obtain $\varphi(z)$, $E(z)$, and then $\Delta\alpha/\alpha(z)$. By
 minimizing the corresponding total $\chi^2$ in Eq.~(\ref{eq16}), we
 find the best-fit model parameters $\Omega_{m0}=0.2786$,
 $\zeta=-0.2161\times 10^{-5}$, $\beta=0_+$, and $v_0=3.5588$,
 while $\chi^2_{min}=863.909$ and $\chi^2_{min}/dof=0.9919$.
 Note that the best-fit $\beta$ is not exactly equal to $0$,
 but it is extremely close to $0$. In Fig.~\ref{fig5}, we also
 present the corresponding $68.3\%$ and $95.4\%$ confidence
 level contours in the $\Omega_{m0}-\zeta$,
 $\Omega_{m0}-\beta$, $\Omega_{m0}-v_0$, $\zeta-\beta$,
 $\zeta-v_0$ and $\beta-v_0$ planes. From Fig.~\ref{fig5}, it
 is easy to see that $\zeta$ is tightly constrained to a
 narrow range of ${\cal O}(10^{-6})$, thanks to the 293
 $\Delta\alpha/\alpha$ data of ${\cal O}(10^{-5})$. On the
 other hand, $\zeta=0$ deviates from the best fit beyond
 $1\sigma$, but is still within the $2\sigma$ region. So,
 the varying $\alpha$ and $\Lambda\propto\alpha^{-6}$ are
 slightly favored by the observational data, while a constant
 $\alpha$ in ordinary $\Lambda$CDM cosmology is
 still consistent with the observational data within the
 $2\sigma$ region. Note that the observational data cannot well
 constrain the parameter $v_0$, while $\beta$ is constrained to
 $\lsim\,{\cal O}(1)$.

%============================= Fig. 5 =================================

%======================================================================

\vspace{-10mm}  % used here just for a comfortable typesetting

%============================= section 4 ===================================

\section{Concluding remarks}\label{sec4}

In the present work, we extend the work of~\cite{r23} by considering
 the mechanism to drive the varying fine-structure ``constant''
 $\alpha$. In~\cite{r23}, the varying $\alpha$
 and $\Lambda\propto\alpha^{-6}$ were studied only in a
 phenomenological manner, from an interacting vacuum energy
 perspective. Instead, here we further consider the scalar
 field $\phi$ coupling with the electromagnetic field, and hence it
 could drive the variation of $\alpha$. The scalar field $\phi$
 is subdominant, and it is used to drive only the varying $\alpha$.
 The accelerated expansion of the universe is instead driven by
 the cosmological constant $\Lambda$ (equivalently the vacuum
 energy). On the other hand, $\Lambda\propto\alpha^{-6}$ was
 derived from three completely independent approaches in the
 literature, especially the so-called axiomatic approach~\cite{r7}.
 So, the two amazing discoveries in 1998 are connected in this
 way. The coupling $B_F(\phi)$ between the scalar field $\phi$
 and the electromagnetic field plays an important role. In this
 work, we consider various forms of the coupling $B_F(\phi)$,
 and confront the varying $\alpha$
 and $\Lambda\propto\alpha^{-6}$ models with the observational
 data. We find that the key model parameter $\zeta$ in the coupling
 $B_F$ can be tightly constrained to the very narrow ranges of
 ${\cal O}(10^{-5})$ or ${\cal O}(10^{-6})$, thanks to the 293
 $\Delta\alpha/\alpha$ data of ${\cal O}(10^{-5})$. In the cases of
 linear, power-law and exponential couplings, a constant $\alpha$ in
 ordinary $\Lambda$CDM cosmology is fully consistent with the
 observational data. There is no evidence for the varying
 $\alpha$ and $\Lambda$. In the case of polynomial coupling,
 the varying $\alpha$ and $\Lambda\propto\alpha^{-6}$ are
 slightly favored beyond $1\sigma$.

%==================== table 1 ====================

 \begin{table}[tb]
 \begin{center}
 \begin{tabular}{lllllllll}\hline\hline\\[-3.5mm]
 ~Model~~~ & IntIcon ~~ & IntICPL ~~ & IntIIcon ~~ & IntIICPL ~ & BFlin ~~~ & BFpl ~~~~~ & BFexp ~~~ & BFpoly~ \\[1.2mm] \hline \\[-3.5mm]
 ~$\chi^2_{min}$ & 868.149 & 856.005 & 870.391 & 857.605 & 869.6 & 868.527 & 869.6 & 863.909 \\
 ~$k$ & 2 & 3 & 2 & 3 & 3 & 3 & 3 & 4 \\
 ~$\chi^2_{min}/dof$ ~~~ & 0.9944 & 0.9817 & 0.9970 & 0.9835 & 0.9972 & 0.9960 & 0.9972 & 0.9919 \\[0.5mm]
 ~$\Delta$BIC & 5.370 & 0 & 7.612 & 1.6 & 13.595 & 12.522 & 13.595 & 14.678 \\
 ~$\Delta$AIC & 10.144 & 0 & 12.386 & 1.6 & 13.595 & 12.522 & 13.595 & 9.904 \\
 ~Rank & 3 & 1 & 4 & 2 & 7 & 6 & 7 & 5 \\[1.2mm] \hline\hline
 \end{tabular}
 \end{center}
 \caption{\label{tab1} Comparing the eight models considered
 in~\cite{r23} and the present work. See the text for details.}
 \end{table}

%=================================================

Some remarks are in order. Firstly, it is worth noting that
 3 of 4 models considered in~\cite{r23} favor the varying
 $\alpha$ and $\Lambda\propto\alpha^{-6}$, while a constant
 $\alpha$ in ordinary $\Lambda$CDM model deviates from the best
 fit beyond $2\sigma$ or at least $1\sigma$. So, the results
 obtained in the present work are quite contrary to the ones
 of~\cite{r23}, while the same observational datasets are used.
 The main difference between this work and~\cite{r23} is that
 different perspectives and then different parameterizations
 are taken. In~\cite{r23}, the varying $\alpha$
 and $\Lambda\propto\alpha^{-6}$ are studied from
 an interacting vacuum energy perspective. Thus,
 the corresponding parameterizations are performed in the
 interaction between the vacuum energy and pressureless
 matter. On the other hand, in this work, the parameterizations
 are performed in the coupling $B_F(\phi)$ between the scalar
 field $\phi$ and the electromagnetic field. So, it is of
 interest to compare the eight models considered in~\cite{r23}
 and the present work. We label the four models characterized
 by Eqs.~(37), (39), (40), (43) of~\cite{r23} as IntIcon,
 IntICPL, IntIIcon, IntIICPL, respectively. We also label
 the four models characterized by the linear, power-law,
 exponential, polynomial couplings in the present work as
 BFlin, BFpl, BFexp, BFpoly, respectively. Since these models
 have different free parameters and the correlations between
 model parameters are fairly different, it is not suitable to
 directly compare their confidence level contours. Instead, it
 is more appropriate to compare them from the viewpoint of
 goodness-of-fit. A conventional criterion for model comparison
 in the literature is $\chi^2_{min}/dof$, in which the degree
 of freedom $dof={\cal N}-k$, while $\cal N$ and $k$ are the
 number of data points and the number of free model parameters,
 respectively. On the other hand, there are other criteria
 for model comparison in the literature. The most sophisticated
 criterion is the Bayesian evidence (see e.g.~\cite{r42} and
 references therein). However, the computation of Bayesian
 evidence usually consumes a large amount of time and power. As
 an alternative, one can consider some approximations of
 Bayesian evidence, such as the so-called Bayesian Information
 Criterion (BIC) and Akaike Information Criterion (AIC). The
 BIC is defined by~\cite{r43}
 \be{eq26}
 {\rm BIC}=-2\ln{\cal L}_{max}+k\ln {\cal N}\,,
 \ee
 where ${\cal L}_{max}$ is the maximum likelihood. In the
 Gaussian cases, $\chi^2_{min}=-2\ln{\cal L}_{max}$. So, the
 difference in BIC between two models is given by
 $\Delta{\rm BIC}=\Delta\chi^2_{min}+\Delta k \ln {\cal N}$.
 The AIC is defined by~\cite{r44}
 \be{eq27}
 {\rm AIC}=-2\ln{\cal L}_{max}+2k\,.
 \ee
 The difference in AIC between two models is
 given by $\Delta{\rm AIC}=\Delta\chi^2_{min}+2\Delta k$. In
 Table~\ref{tab1}, we present $\chi^2_{min}/dof$, $\Delta$BIC
 and $\Delta$AIC for the eight models considered in~\cite{r23}
 and the present work. Note that the IntICPL model has been
 chosen to be the fiducial model when we calculate $\Delta$BIC
 and $\Delta$AIC. Clearly, all the four models considered
 in~\cite{r23} are better than all the four models considered
 in the present work. The IntICPL model is the best from the
 viewpoint of all the three criteria $\chi^2_{min}/dof$, BIC
 and AIC.

Secondly, in this work we consider the scalar field $\phi$
 without potential for simplicity. In general, the potential
 could be included. However, including the potential will
 significantly increase the model's degree of freedom, since
 the forms of potential can be diverse, and the number of model
 parameters used to define the potential might be large. All
 these will make the constraints fairly loose. Similarly, one
 can consider the other complicated scalar fields, such as
 $k$-essence, Dirac-Born-Infeld scalar field, tachyon, in
 place of quintessence considered in this work. But this will
 significantly increase the model's degree of freedom, and
 greatly decrease the constraining ability.

Thirdly, from Figs.~\ref{fig2}--\ref{fig5}, it is interesting
 to find that the contours in the $\Omega_{m0}-\zeta$ plane
 are nearly symmetric and their shapes are close to circles
 (we thank the referee for pointing out this issue). In
 fact, this shows that the coupling constant $\zeta$ is
 largely uncorrelated with $\Omega_{m0}$. Thus, it further
 justifies our parameterizations (\ref{eq17}), (\ref{eq21}),
 (\ref{eq22}) and (\ref{eq24}) for the coupling $B_F$.

Finally, we again advocate the idea of $\Lambda(\alpha)$CDM
 cosmology with $\Lambda\propto\alpha^{-6}$ while
 the fine-structure ``constant'' $\alpha$ is varying. In fact,
 although $\Lambda\propto\alpha^{-6}$ could be derived from
 various completely independent approaches, it has not
 attracted considerable attention in the community so far.
 But it is impressive that the numerical value of
 $\Lambda=G^2 m_e^6/(\hbar^4 \alpha^6)$ from Eq.~(\ref{eq1})
 is very close to the observational value. There might be a
 profound reasoning, other than just a coincidence. On the
 other hand, if $\alpha$ is varying, the well motivated
 $\Lambda(\alpha)\propto\alpha^{-6}$ gives a novel realization
 of $\Lambda(t)$, different from the ones purely written by
 hand in the literature. We consider that it deserves further
 investigation.

%============================= acknowledgements ===================================

\section*{ACKNOWLEDGEMENTS}

We thank the editors and the anonymous referee for quite useful
 comments and suggestions, which helped us to improve this work. We
 are grateful to Profs.~Rong-Gen~Cai and Shuang~Nan~Zhang for
 helpful discussions. We also thank Minzi~Feng, as well as
 Zu-Cheng~Chen, Shou-Long~Li, Xiao-Bo~Zou,  Hua-Kai~Deng and
 Zhao-Yu~Yin for kind help and discussions. This work was supported
 in part by NSFC under Grants No.~11575022 and No.~11175016.

\renewcommand{\baselinestretch}{1.0}

%============================= references ==================================

\end{document}